\documentclass[twoside,slac_one]{revtex4}
\usepackage{graphicx}
\usepackage{fancyhdr}
\usepackage{amsmath} % American Mathematics Society standards
\usepackage{bm}% bold math
\usepackage{amsxtra}
\usepackage{amssymb}
\usepackage{amsthm}
\usepackage{latexsym}
\usepackage{lscape}

\usepackage{subfig}

\begin{document}

\title{Measurements of the forward energy flow and forward jet production with CMS}


\author{H. Van Haevermaet, for the CMS Collaboration}
\affiliation{Department of Physics, University of Antwerp, Antwerp, Belgium}

\begin{abstract}
We present measurements of the forward ($3 < |\eta| < 5$) energy flow in minimum bias events and in events with either hard jets or W and Z bosons produced at central rapidities together with first measurements of the inclusive forward jet cross section and central forward jet correlations.  Results are compared to MC models with different parameter tunes for the description of the underlying event. 
\end{abstract}

\maketitle

\thispagestyle{fancy}


%%%%%%%%%%%%%%%%%%%%%%%%%%%%%%%%%%
\section{Introduction}
\begin{figure}[h]
\centering
\includegraphics[width=50mm]{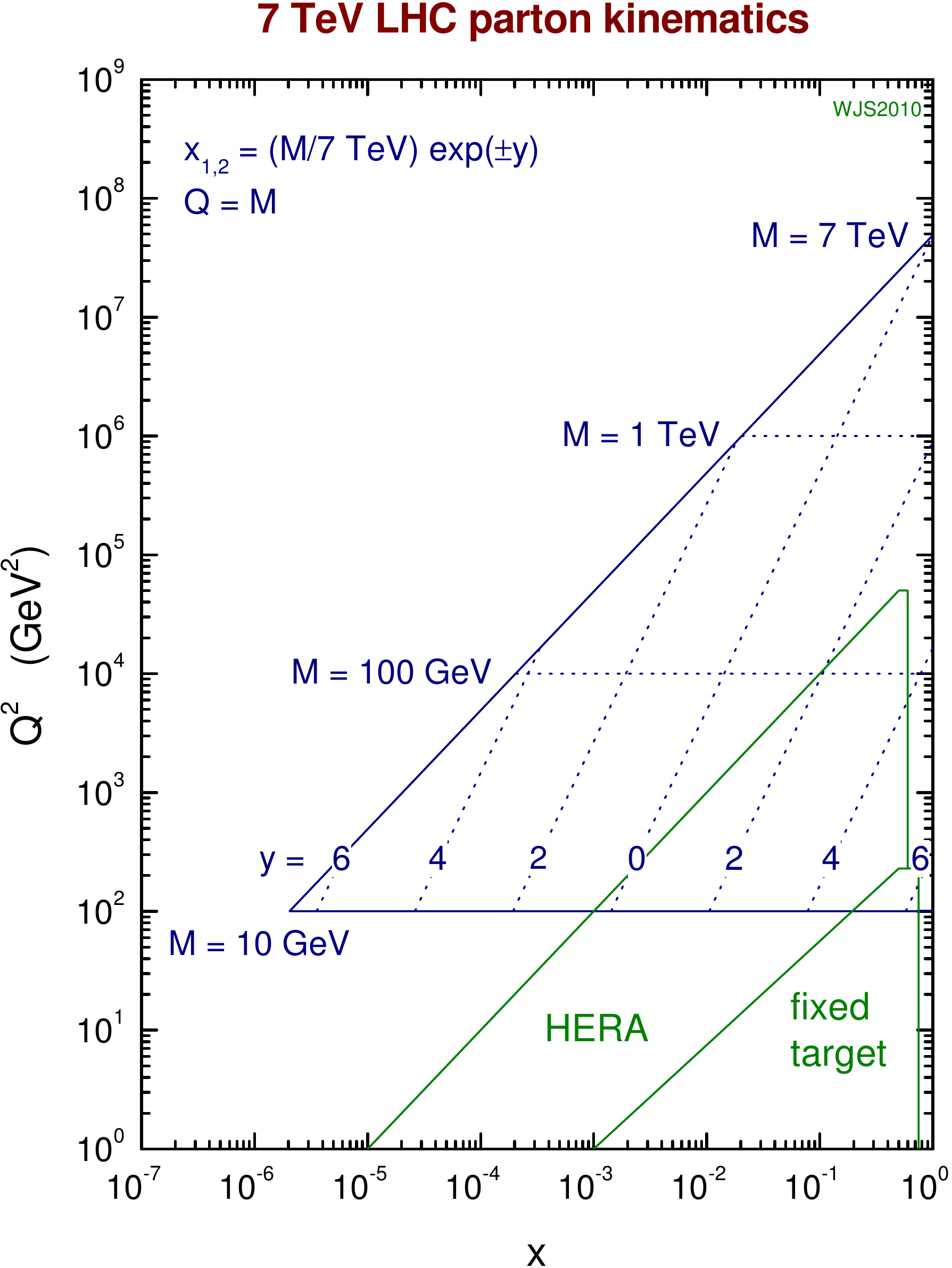}
\caption{LHC parton kinematics ($Q^{2}$,$x$) phase space at $\sqrt{s}$ = 7 TeV compared to HERA. \cite{partonkinematics} } \label{LHCparton}
\end{figure}

The Large Hadron Collider at CERN, which currently delivers proton-proton collisions at $\sqrt{s}$ = 7 TeV, is very suitable to study forward physics since it is basically a small $x$ QCD machine. Figure \ref{LHCparton} shows that the available ($Q^{2},x$) phase space at 7 TeV, with $Q^{2}$ the hard scale of the collision and $x$ the momentum fraction of the proton carried by the struck parton, already extends to values of $x \sim 10^{-5}$ when going to pseudorapidity $\eta \sim 5$. 
Measurements in this forward region enable to probe the content of the proton at small values of the momentum fraction $x$ carried by the partons, in a region where the parton densities might become very large, the probability for more than one partonic interaction per event should increase and the appearance of novel QCD dynamics is expected.

% energy flow intro
A measurement of the forward energy flow is sensitive to parton shower dynamics and the Underlying Event activity (UE) and is thus complementary to those performed in the central region. The measurement therefore enables to study the multiple parton interactions (MPI), which are at the origin of the UE activity. Within the Pythia framework, the MPI are strongly correlated with the impact parameter and the $p_{T}$ of the hard scattering, hence a measurement of the energy flow in Minimum Bias events and events with a hard scale (jets or W/Z bosons) can constrain the existing MPI and UE models. \cite{HFeflow}

% inclusive jet cross section + forward-central intro
The inclusive forward jet cross section is sensitive to parton radiation, underlying partonic QCD processes and to the parton density functions (PDF)  of the incoming hadrons. Previous measurements show that jet cross sections can be described by perturbative QCD over several orders of magnitude in $x$ and $Q^{2}$. They are however limited to central pseudorapidities where the momentum fractions $x_{1}$ and $x_{2}$ of the incoming partons are of the same order. Forward/backward jets however result from scatterings between partons where $x_{2} \ll x_{1}$ and allow one to study QCD in the small $x$ region where the PDF's are less constrained and deviations from DGLAP parton dynamics (e.g. BFKL, CCFM) or gluon saturation effects are expected to appear. A measurement of the simultaneous production of forward and central jets gives additional information on MPI and allows a study of multi-jet production, but in particular detailed studies of different types of parton radiation dynamics like e.g. DGLAP, BFKL, CCFM are possible. \cite{inclusivejetcrosssection}\cite{forwardcentraljets}

% VBF
Jet measurements in the forward region are also of interest for vector-boson-fusion processes like e.g. the production of a Higgs boson with forward and backward jets at $p_{T} \approx$ 40 GeV/c since it is essential to understand the dynamics of forward jet production to control the backgrounds.

%%%%%%%%%%%%%%%%%%%%%%%%%%%%%%%%%%
\section{The CMS detector}
\begin{figure}[h]
\centering
\includegraphics[width=90mm]{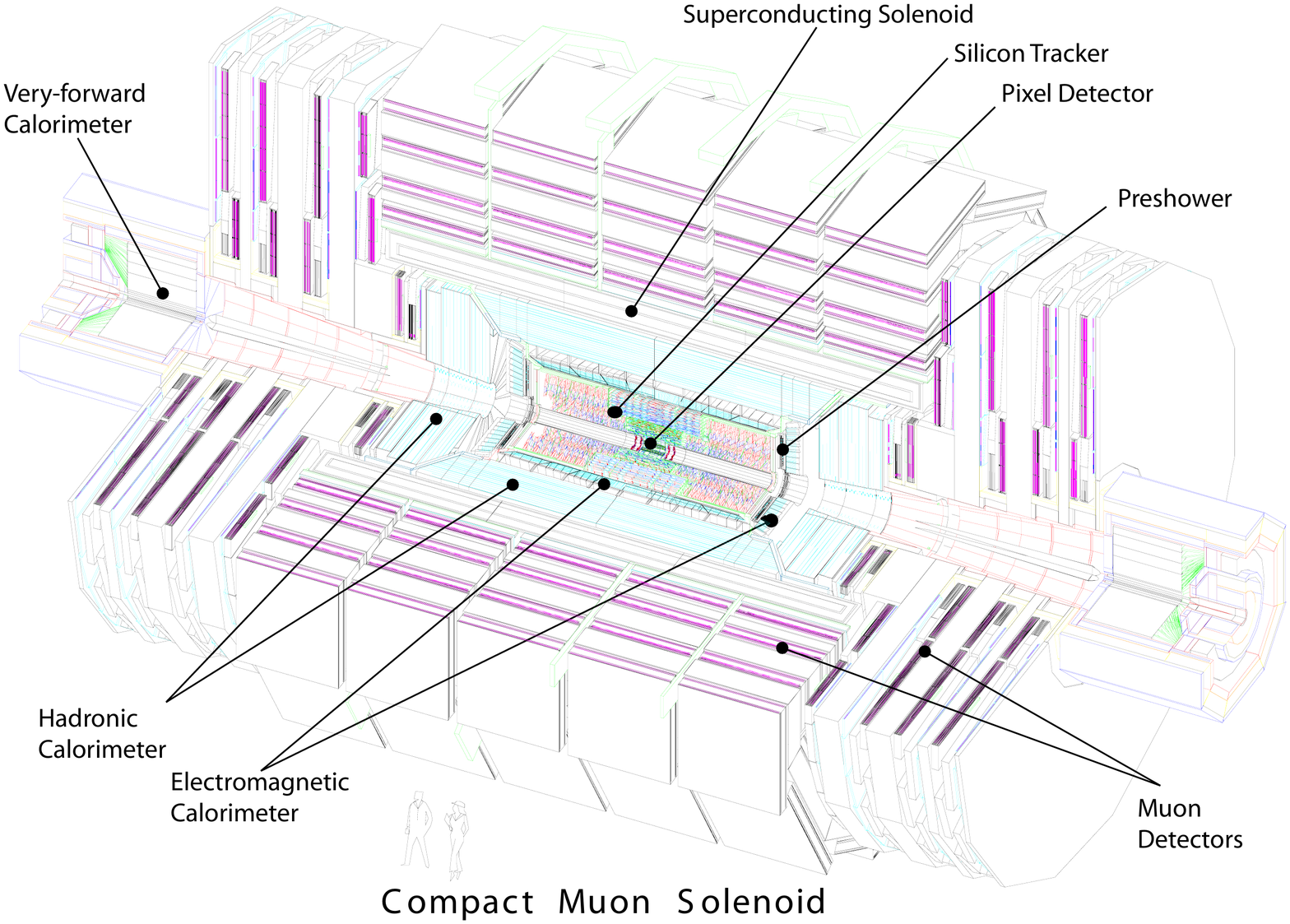}
\caption{Labelled drawing of the CMS detector. The position of one of the HF detectors is indicated with Very-forward calorimeter. \cite{CMSdrawing}} \label{CMS}
\end{figure}

% CMS
A detailed description of the CMS detector can be found elsewhere \cite{CMSpaper}, but the key components in the central region used for the event selections are: a superconducting solenoid generating a 3.8 T magnetic field with inside a silicon pixel detector, strip tracker, crystal electromagnetic calorimeter and a brass-scintillator hadronic calorimeter. Outside the field volume are gaseous muon detectors embedded in iron return yokes. A key component in measuring the forward energy flow and forward jets in these analyses are the Hadronic Forward (HF) calorimeters which are located at $\pm$ 11.2 meters from the interaction point and have a pseudorapidity range of 2.9 $<$ $|\eta|$ $<$ 5.2 (figure \ref{CMS}).  They use iron absorbers and embedded radiation hard quartz fibers to collect the Cherenkov light detected by radiation hard photomultiplier tubes. Both HF calorimeters are segmented in 13 rings in $\eta$. To trigger the event readout two systems are used. The Beam Pick-up Timing eXperiment (BPTX) devices were used to get information on the bunch structure and timing of the beams while the Beam Scintillator Counters (BSC), located in front of HF, were used to detect activity in the pseudorapidity range 3.23 $< |\eta| <$ 4.65. \cite{WZeflow}

%%%%%%%%%%%%%%%%%%%%%%%%%%%%%%%%%%
\section{Analysis strategies}

%%%%%%%%%%%%%%%%%%%%%%%%%%%%%%%%%%
\subsection{Measurement of the forward energy flow}
\begin{figure}[h]
  \centering
    \subfloat[]{\label{fig:minbias}
		\includegraphics[width=35mm]{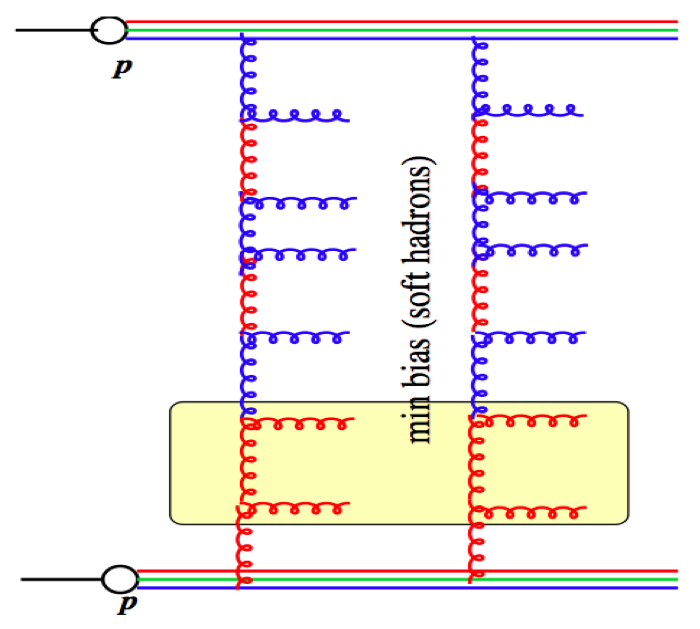}}
	\quad
	  \subfloat[]{\label{fig:dijets}
		\includegraphics[width=37mm]{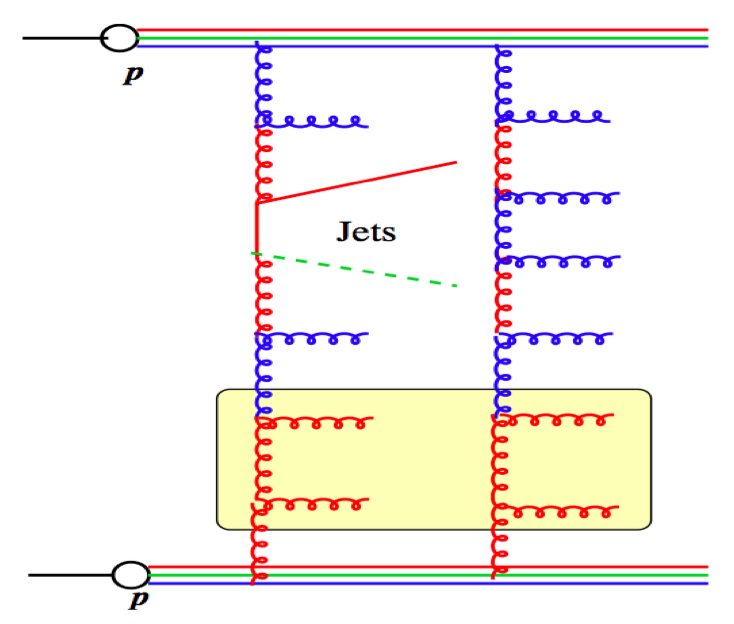}}
	\quad
	  \subfloat[]{\label{fig:WZ}
		\includegraphics[width=31mm]{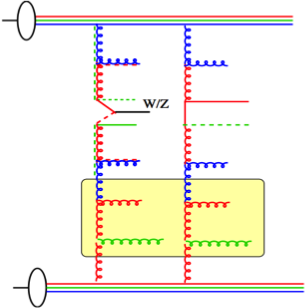}}
\caption{ (a) Diagram of a typical minimum bias event. (b) Diagram of an event with a hard scale present in the form of jets. (c) Diagram of an event with a centrally produced W/Z. \cite{diagrams}}
\end{figure}

% event classes and forward energy flow
The forward energy flow measurements presented here are determined for 3 different event classes: minimum bias events (figure \ref{fig:minbias}) and events with a hard scale provided by a di-jet system in the central region (figure \ref{fig:dijets}) or by a centrally produced W/Z boson (figure \ref{fig:WZ}). The forward energy flow itself is the total energy measured by the HF calorimeters by using the sum of all energy deposits of HF towers above noise threshold (4 GeV). Due to a not fully sufficient simulation of the first 2 and last rings only 10 $\eta$ rings are eventually used in the analysis resulting in an actual forward energy flow measurement in the pseudorapidity range 3.15  $< |\eta| <$ 4.9. 

% event selection
The analyses use low luminosity LHC $pp$ data at $\sqrt{s}$ = 900 GeV and 7 TeV collected in 2010. Basic minimum bias event selection is done using different criteria. Events are required to have a trigger signal in both BSC detectors which coincides with a signal in both BPTX detectors. The former condition ensures a sample close to non single diffraction while the latter ensures the presence of both beams passing the interaction point. Further event cleaning is done by asking at least one good primary reconstructed vertex and beam-induced background events, characterized by an anomalous large number of pixel hits, are filtered out by requiring that if an event has more than 10 tracks, 25$\%$  of all the reconstructed tracks have to be of high purity. 

% dijet selection
On top of the common minimum bias selection, the dijet events are required to have two well reconstructed jets in the pseudorapidity region $|\eta| < $ 2.5 with a $p_{T} >$ 8 GeV at  $\sqrt{s}$ = 900 GeV and $p_{T} >$ 20 GeV at  $\sqrt{s}$ = 7 TeV. Additionally a back to back condition is applied asking for $|\Delta\phi(jet_{1},jet_{2}) - \pi| < 1.0$. The default jet algorithm used is anti-$k_{T}$ with R = 0.5.

% W/Z selection
The W/Z events are selected according to a dedicated procedure corresponding to the identification of $WX \rightarrow l\nu X$ and $ZX \rightarrow ll X$ events. The first type of events are required to have an isolated electron or muon with $p_{T} >$ 25 GeV and $|\eta| <$ 1.4. Events containing a second electron or muon with $p_{T} >$ 10 GeV are rejected and the missing transverse momentum of the neutrino candidate must be bigger than 30 GeV while the transverse mass of the lepton and neutrino are required to be larger than 60 GeV. The Z decay type of events are required to have two isolated electrons or muons with opposite charge, $p_{T} >$ 25 GeV and at least one of them must have a pseudorapidity $|\eta| <$ 1.4. Furthermore, the invariant mass of the di-lepton pair must be between 60 and 120 GeV. \cite{HFeflow} \cite{WZeflow}

%%%%%%%%%%%%%%%%%%%%%%%%%%%%%%%%%%
\subsection{Forward jet production measurements}
% inclusive forward jets
The inclusive forward jet cross section measurement presented here is based on LHC $pp$ collisions recorded in 2010 at $\sqrt{s}$ = 7 TeV. An uncorrected single-jet trigger with a $p_{T}$ threshold of 15 GeV/c within $|\eta| <$ 5.0 is used. The final sample corresponds to an integrated luminosity of 3.14 pb$^{-1}$. All events are required to have a good reconstructed primary vertex and beam-induced background events are filtered out. Furthermore, events with anomalous noise in the HF calorimeters are also removed from the analysis. 

\begin{figure}[h]
  \centering
    \subfloat[]{\label{fig:forwardjet}
		\includegraphics[width=30mm]{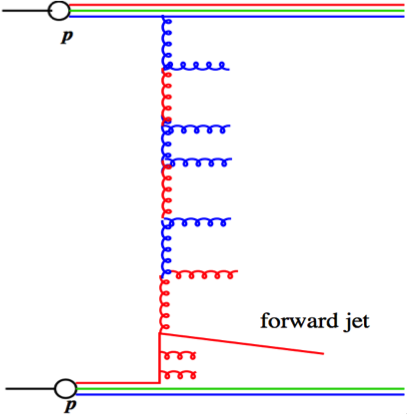}}
	\quad
	  \subfloat[]{\label{fig:forwardcentraljets}
		\includegraphics[width=30mm]{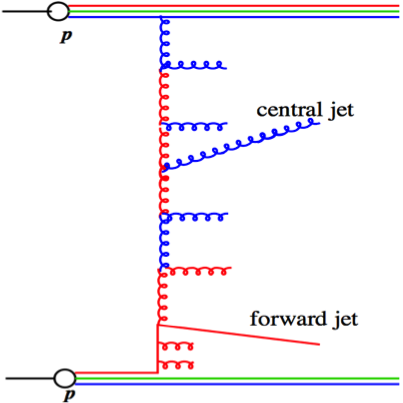}}
\caption{ (a) Diagram of an event with a forward jet. (b) Diagram of simultaneous central and forward jet production. \cite{diagrams}}
\end{figure}

Jets are reconstructed using the anti-$k_{T}$ algorithm with R = 0.5 and are required to have their axis between 3.2 $< |\eta| <$ 4.7. The transverse momentum of the jets must be larger than 20 GeV and various jet quality criteria are applied to remove fake jets. The efficiency of the above mentioned single-jet trigger is measured using another single jet trigger with lower threshold ($p_{T} >$ 6 GeV) and is found to be fully efficient when the corrected $p_{T}$ is above 35 GeV/c. This value is taken as the lower limit for the cross section measurement. After all selections, only events with at least one forward jet (figure \ref{fig:forwardjet}) remain. 

\newpage
% central-forward selection
For the cross section measurement of simultaneous production of a central and forward jet (figure \ref{fig:forwardcentraljets}) the same data sample taken in 2010 at $\sqrt{s}$ = 7 TeV with an integrated luminosity of  3.14 pb$^{-1}$ is used. Online event selection is done using a di-jet trigger with an uncorrected calorimeter energy threshold of $(E_{T,1}+E_{T,2})/2 >$ 15 GeV where the $E_{T}$ is integrated within $|\eta| <$ 5.2. Analog to the previous event selections, events are required to have at least one good reconstructed primary vertex and beam-induced background events are filtered. Jet are reconstructed with the anti-$k_{T}$ (R = 0.5) algorithm. The central jet is defined to be within $|\eta| <$ 2.8 while the forward jet within 3.2 $< |\eta| <$ 4.7 and an event is accepted when at least one central and one forward jet with both $p_{T} >$ 35 GeV are present. The efficiency of the high level di-jet trigger has been studied using a minimum bias sample and is found to be 100$\%$ for jets with a $p_{T} >$ 35 GeV. \cite{inclusivejetcrosssection} \cite{forwardcentraljets}

%%%%%%%%%%%%%%%%%%%%%%%%%%%%%%%%%%
\section{Hadron level corrections}

% hadron level correction done
All measurements, except the forward energy flow analysis with W/Z events, are corrected to hadron level using bin-by-bin methods which take into account acceptances, inefficiencies and bin migrations due to the finite detector resolution. The correction factors are defined as ratios of MC predictions at hadron level and detector level where the detector level MC is defined as a fully simulated and reconstructed sample which goes through the same analysis chain as data. These factors are calculated for different MC tunes and the average value is taken for the final correction. The deviation due to the different models is included as a model dependent systematic error.

% forward energy flow def on hadron level
The forward energy flow on hadron level is defined by summing the energy of final stable particles in HF (3.15 $< |\eta| <$ 4.9) and excluding neutrinos and muons. In order to keep the correction applied low, diffractive events on hadron level are suppressed using a selection similar to the BSC trigger used in data. The kinematic cuts to select the di-jet events are the same as the ones used in data.

% jets on hadron level
Before the forward jet spectrum can be corrected to hadron level, the jet energy scale and resolution need to be corrected. The jet energy scale corrections, derived from data and simulations, take into account the $\eta$ and $p_{T}$ dependence of the HF calorimeters response in order to determine the absolute jet energy calibration. The relative energy resolution is found to be below $\sim 12\%$ when $p_{T} >$ 35 GeV/c and the jet position resolution is $\sigma_{\phi,\eta} \approx 0.035$ at $p_{T} =$ 20 GeV and improves to $\sigma_{\phi,\eta} \approx 0.02$ for $p_{T} >$ 100 GeV. Hence the binning of the presented distributions is equal or larger than the experimental resolutions. Finally a bin-by-bin correction is applied to go to hadron level by using simulated samples to study the $p_{T}$ bin migrations in terms of acceptance, background, purity and stability. \cite{HFeflow} \cite{inclusivejetcrosssection} \cite{forwardcentraljets}

\section{Systematics uncertainties}
In this section the main systematic uncertainties present in the different measurements are listed. Table \ref{table:HFeflow_systematics} shows the three dominant contributions to the forward energy flow measurement in HF. All other effects are found to be less then 3$\%$. 
\begin{table}[h]
\begin{center}
\caption{Three main contributions to the systematic uncertainty on the forward energy flow in HF. \cite{HFeflow}}
\begin{tabular}{|l|c|c|}
\hline \textbf{Uncertainty} & \textbf{on minimum bias} & \textbf{on di-jet}
\\
\hline Energy scale & 10$\%$ & 10$\%$ \\
\hline Simulation (short fibre response), $\sqrt{s} =$ 0.9/7 TeV & 3-9$\%$/3-6$\%$ & 6-18$\%$/6-8$\%$  \\
\hline Model dependence, $\sqrt{s} =$ 0.9/7 TeV & 1-3$\%$/1-2$\%$ & 4-11$\%$/12-17$\%$ \\
\hline
\end{tabular}
\label{table:HFeflow_systematics}
\end{center}
\end{table}

% systematics inclusive forward jets
The dominating systematic uncertainties on the forward jet cross section measurements are the jet energy scale, resolution and the integrated luminosity. The jet energy calibration uncertainty varies between 3-6$\%$ and propagates to 20-30$\%$ on the final cross section. The jet $p_{T}$ resolution of 10$\%$ results in an uncertainty of 3-6$\%$ and the integrated luminosity gives an additional uncertainty of 4$\%$ on the final cross section. Two smaller contributions, the presence of pile-up and the model uncertainty of the bin-by-bin corrections, give additional uncertainties of 5$\%$ and 3$\%$ respectively. \cite{HFeflow} \cite{inclusivejetcrosssection} \cite{forwardcentraljets}
%%%%%%%%%%%%%%%%%%%%%%%%%%%%%%%%%%
\section{Results}
\subsection{Measurement of the forward energy flow}
\begin{figure}[h]
  \centering
    \subfloat[]{\label{fig:eflow_minbias}
		\includegraphics[width=70mm]{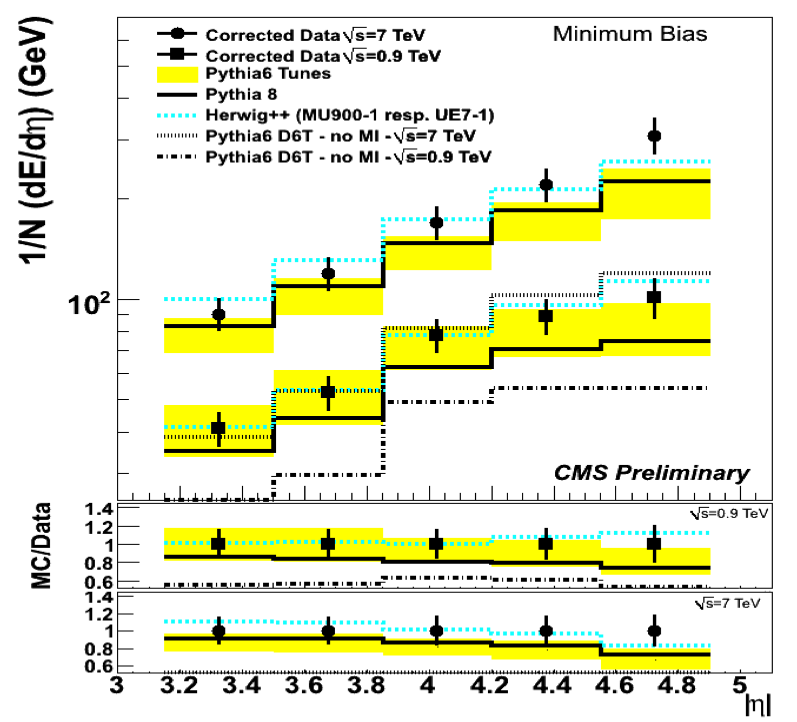}}
	\quad
	  \subfloat[]{\label{fig:eflow_minbias_CR}
		\includegraphics[width=70mm]{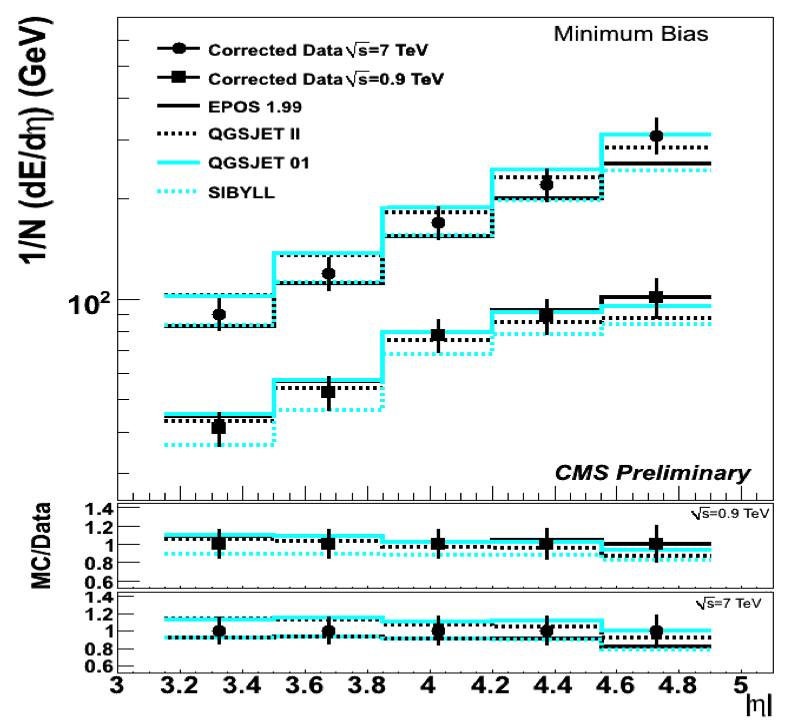}}
\caption{ (a) Energy flow in HF for minimum bias events at $\sqrt{s}$ = 900 GeV and $\sqrt{s}$ = 7 TeV compared to various Pythia models and Herwig++. (b) Energy flow in HF for minimum bias events at $\sqrt{s}$ = 900 GeV and $\sqrt{s}$ = 7 TeV compared to $pp$ generators used in cosmic rays physics. \cite{HFeflow}}
\end{figure}

Figure \ref{fig:eflow_minbias} shows the corrected forward energy flow (3.15  $< |\eta| <$ 4.9) in minimum bias events for $\sqrt{s}$ = 900 GeV and  $\sqrt{s}$ = 7 TeV compared to several Pythia models and Herwig++. The spread of the different Pythia6 tunes (CW, D6T, DW, ProQ20, Z2, P0 and ProPT0) is indicated by a yellow band while the results Pythia6 D6T tune without MPI are plotted separately. The observed rise with $\eta$ corresponds to a flat $E_{T}$ flow and the increase from 900 GeV to 7 TeV is similar to the increase of the number of charged particles. It is clear that predictions of models without MPI are too low and a large spread between the different Pythia6 tunes is observed while Herwig++ seems to work reasonably well. Figure  \ref{fig:eflow_minbias_CR} shows the same corrected energy flow measurement, but compared to various  $pp$ generators used in Cosmic Ray physics. It appears that these work quite well without additional tuning. 

\begin{figure}[h]
  \centering
    \subfloat[]{\label{fig:eflow_dijets_7TeV}
		\includegraphics[width=69mm]{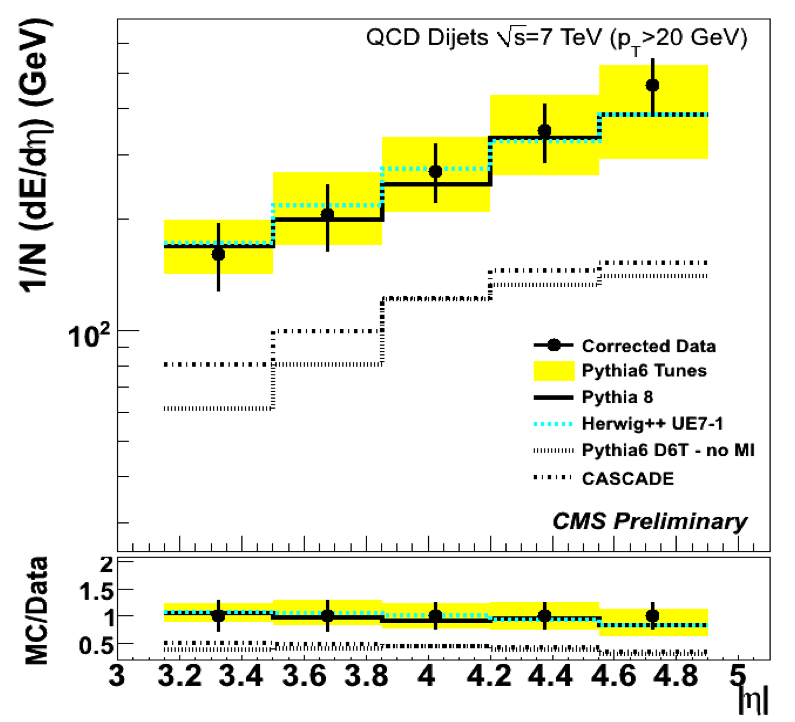}}
	\quad
	  \subfloat[]{\label{fig:eflow_dijets_7TeV_CR}
		\includegraphics[width=69mm]{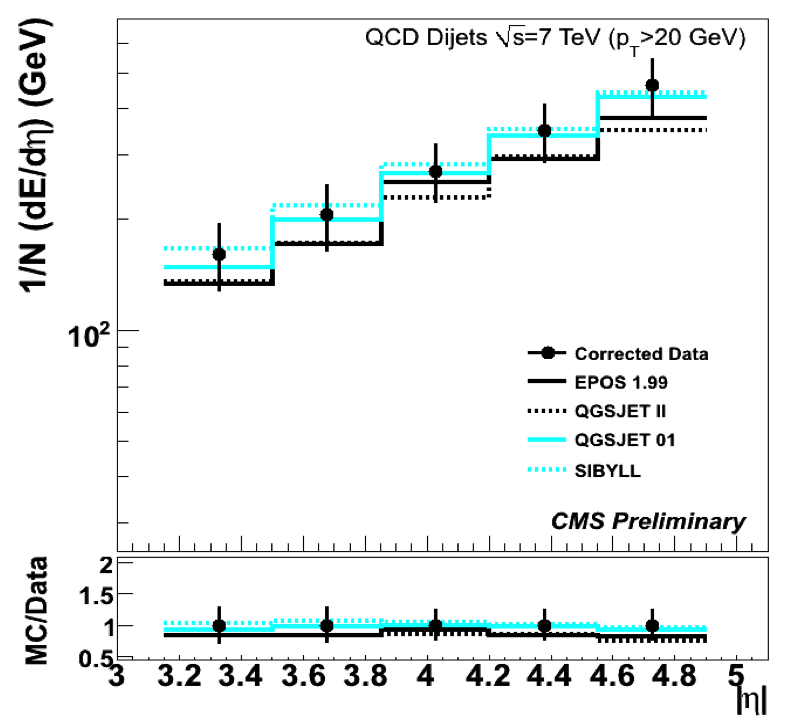}}
\caption{ (a) Energy flow in HF for di-jet events at $\sqrt{s}$ = 7 TeV compared to various Pythia models, Herwig++ and CASCADE.  (b) Energy flow in HF for dijet events at $\sqrt{s}$ = 7 TeV compared to $pp$ generators used in cosmic rays physics. \cite{HFeflow}}
\end{figure}

Figure \ref{fig:eflow_dijets_7TeV} shows the corrected forward energy flow at  $\sqrt{s}$ = 7 TeV for the selected di-jet events compared to Pythia models, Herwig++ and CASCADE. The energy flow of the Pythia6 D6T tune without MPI is still too low. However the different Pythia models with MPI cover the data while CASCADE predicts a much smaller energy flow. Figure \ref{fig:eflow_dijets_7TeV_CR} shows the same corrected di-jet energy flow at  $\sqrt{s}$ = 7 TeV compared to $pp$ generators used in Cosmic Ray physics.

\begin{figure}[h]
\centering
\includegraphics[width=78mm]{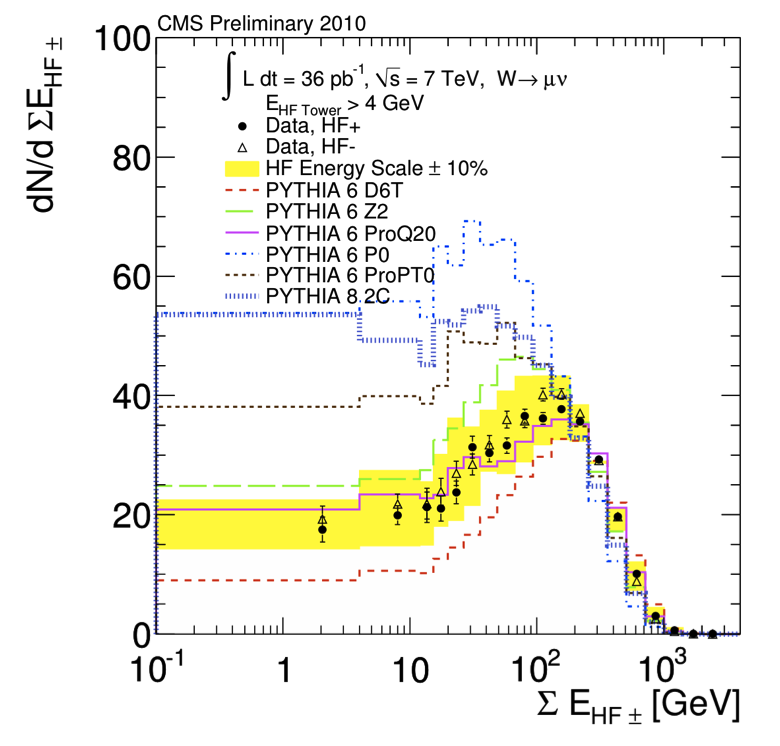}
\caption{Energy flow in the HF+ and HF- detectors for events with a central W boson compared to various Pythia models. \cite{WZeflow} } \label{fig:WZenergyflow}
\end{figure}

% W/Z results
The measured forward energy flow distribution at detector level in HF in events with a centrally produced W boson is shown in figure \ref{fig:WZenergyflow}. It compares the total recorded energy in HF$_{\pm}$ with different Pythia models. The yellow band indicates the HF energy scale uncertainty of 10$\%$. The distribution, which is sensitive to different UE tunes, shows clearly large differences at lower and higher energy regions depending on which model is used. \cite{HFeflow} \cite{WZeflow}

\subsection{Forward jet production measurements}
\begin{figure}[h]
\centering
\includegraphics[width=90mm]{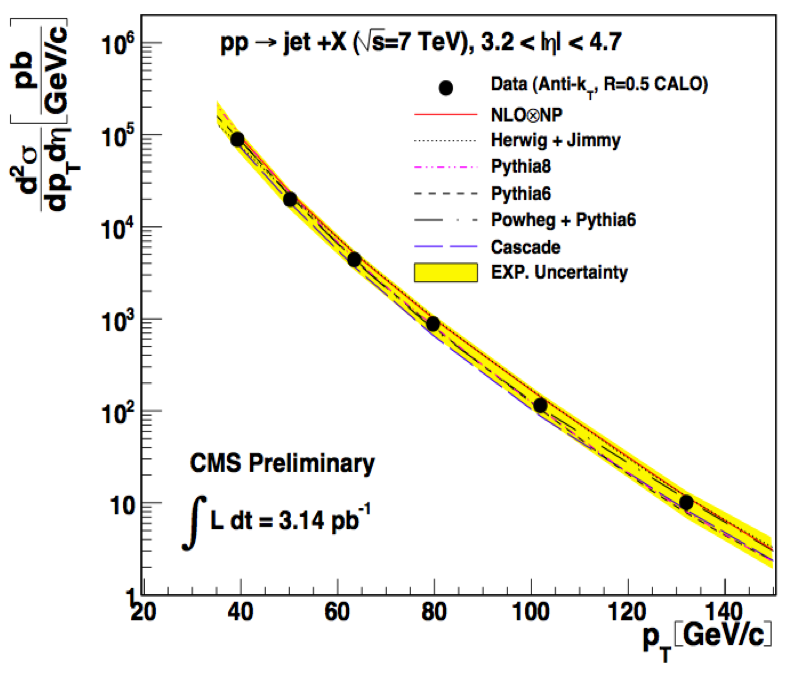}
\caption{The measured inclusive forward jet cross-section at $\sqrt{s}$ = 7 TeV compared to various Pythia models, Herwig+Jimmy, Powheg, CASCADE and NLO calculations.  \cite{inclusivejetcrosssection}} \label{fig:inclusiveforwardjets}
\end{figure}

\begin{figure}[h]
  \centering
    \subfloat[]{\label{fig:forwardcentraljets_cross_section_1}
		\includegraphics[width=71mm]{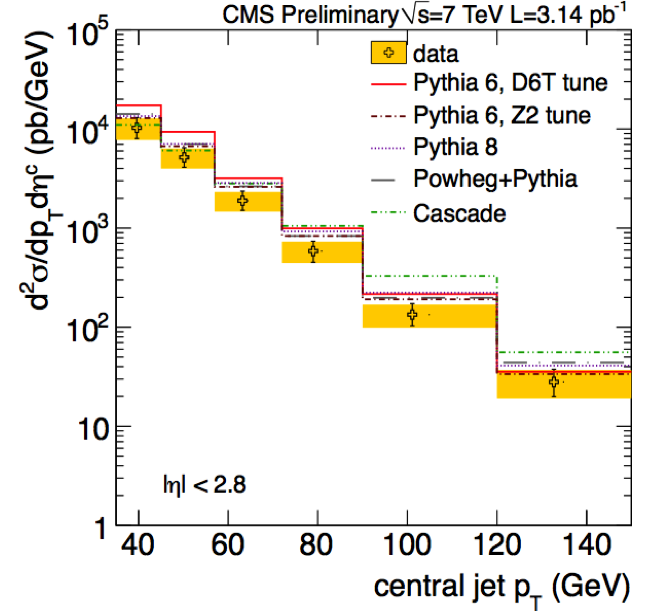}}
	\quad
	  \subfloat[]{\label{fig:forwardcentraljets_cross_section_2}
		\includegraphics[width=70mm]{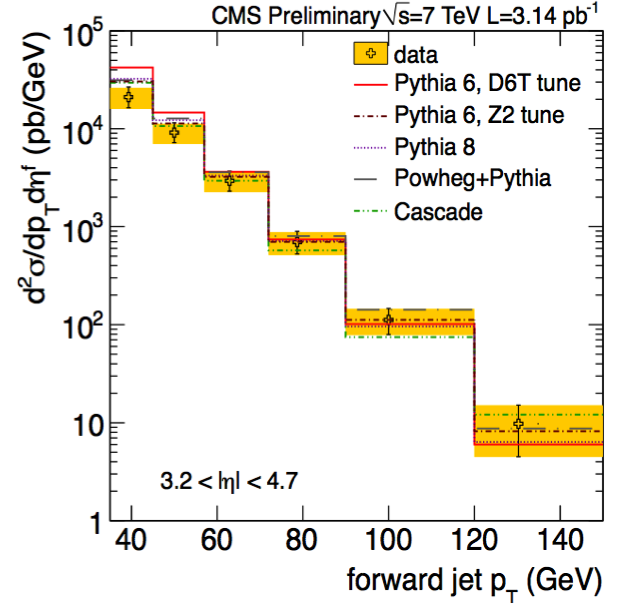}}
	\quad
	  \subfloat[]{\label{fig:forwardcentraljets_cross_section_4}
		\includegraphics[width=70mm]{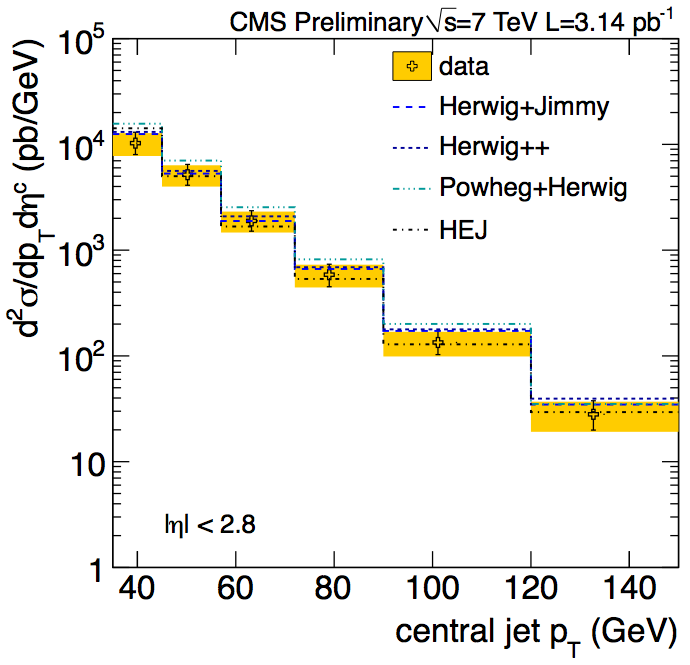}}
	\quad
	  \subfloat[]{\label{fig:forwardcentraljets_cross_section_3}
		\includegraphics[width=70mm]{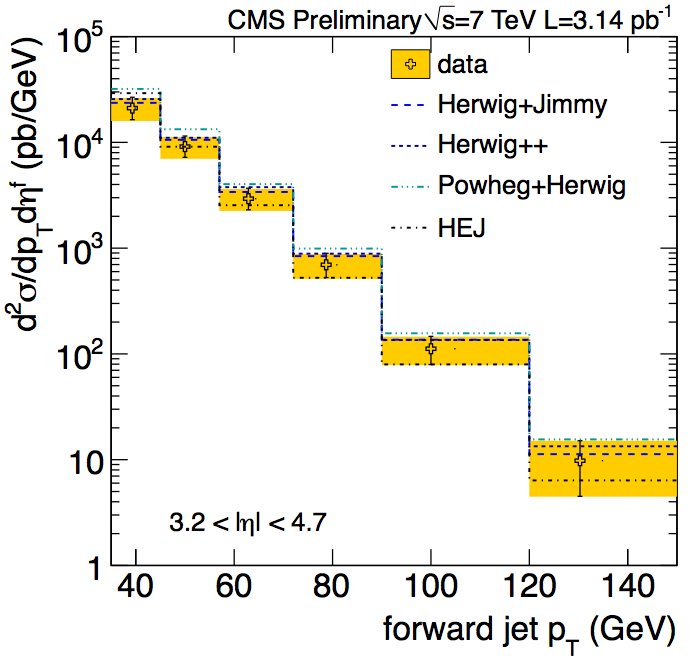}}
\caption{ Measured differential cross section at $\sqrt{s}$ = 7 TeV for the simultaneous production of a central and forward jet versus the central jet $p_{T}$ (a)(c) and versus the forward jet $p_{T}$ (b)(d) compared to different hadron level models. \cite{forwardcentraljets} }
\end{figure}

The corrected cross section measurements of the inclusive forward jet production and simultaneous production of a central and forward jet are shown in figures \ref{fig:inclusiveforwardjets}, \ref{fig:forwardcentraljets_cross_section_1} and \ref{fig:forwardcentraljets_cross_section_2}. 
The plotted differential cross section is defined as:
\begin{equation}
\frac{d^{2}\sigma}{dp_{T}d\eta} = \frac{C_{unfold}}{\cal{L}}\frac{N_{jets}}{\Delta p_{T}\Delta \eta} [pb/(GeV/c)]
\label{eq-crosssection}
\end{equation}
where $N_{jets}$ is the number of jets measured in a particular $p_{T}$ and $\eta$ bin with widths $\Delta p_{T}$ and $\Delta \eta$, $\cal{L}$ = 3.14 pb$^{-1}$ the total integrated luminosity and $C_{unfold}$ is the detector to hadron level correction factor. Figure \ref{fig:inclusiveforwardjets} shows the inclusive forward jet cross section at pseudorapidities 3.2 $< |\eta| <$ 4.7 compared to several hadron level models (NLO calculations, POWHEG, Pythia 6, Pythia 8, HERWIG 6 and CASCADE). The yellow band indicates the total systematic uncertainty which is dominated by the jet energy scale. Within current experimental and theoretical uncertainties all calculations reproduce the measured forward jet cross section for 35 GeV $< p_{T} <$ 150 GeV/c. 

Figures \ref{fig:forwardcentraljets_cross_section_1}, \ref{fig:forwardcentraljets_cross_section_4} and \ref{fig:forwardcentraljets_cross_section_2}, \ref{fig:forwardcentraljets_cross_section_3} show the cross sections for simultaneous production of a central and forward jet versus the central jet $p_{T}$ and forward jet $p_{T}$. The band around the data points indicates the total systematic uncertainty, dominated by the jet energy scale. 
The measurement is compared to two Pythia 6 tunes (D6T and Z2), Pythia 8, POWHEG, CASCADE, HERWIG and HEJ. The Pythia tunes overestimate the jet spectra at low $p_{T}$ values. NLO contributions, which are included in POWHEG, do not improve the description significantly and CASCADE has large deviations with respect to the data. HERWIG and HEJ however give a better description of the data. \cite{inclusivejetcrosssection} \cite{forwardcentraljets}

\section{Conclusions}
We presented the first measurements done at low and high $p_{T}$ in the forward region at CMS for data at $\sqrt{s}$ = 900 GeV and  $\sqrt{s}$ = 7 TeV. In a range of 3.15  $< |\eta| <$ 4.9 the energy flow has been measured for minimum bias events, di-jet events and W/Z boson events. It is shown that the inclusion of MPI in the models improves the description of data and that none of the used MC models can describe all energy flow measurements. Furthermore it is noted that Cosmic Ray models work quite well. The large spread of the different Pythia tunes illustrates that the forward measurement is complementary to the central Underlying Event studies to improve the MC generators. 
%\newpage
A first measurement of the inclusive forward jet cross section has been done in the region 3.2 $< |\eta| <$ 4.7 and provides a first test of perturbative QCD calculations in this forward region at high energies. Within the current experimental uncertainties however, which are dominated by the jet energy scale, all models are able to describe the data.

A measurement of the cross section of simultaneous central and forward jet production has been presented. It gives more information on MPI and allows to study different types of parton radiation dynamics. Even with the current large experimental uncertainties, the measured correlations between the forward and central regions already show that the theory does not describe the data in a satisfactory way.

%\begin{acknowledgments}
%This document is adapted from the instructions provided to the authors
%of the proceedings papers at CHARM~07, Ithaca, NY,  
%and from eConf templates.
%\end{acknowledgments}

\bigskip % extra skip inserted

\end{document}